\documentstyle[epsfig,eqsecnum,aps]{revtex}
\begin{document}
\draft
\title{Isospin Mixing and Fermi Transitions: Selfconsistent Deformed 
Mean Field Calculations and Beyond}
\author{R. \'Alvarez-Rodr\'{\i}guez, E. Moya de Guerra and P. Sarriguren} 
\address{Instituto de Estructura de la Materia,
Consejo Superior de Investigaciones Cient\'{\i }ficas, \\
Serrano 123, E-28006 Madrid, Spain}
%\date{\today}
\maketitle

\begin{abstract}

We study Fermi transitions  and isospin mixing in an isotopic chain 
($^{70-78}$Kr) considering various approximations that use the same 
Skyrme-Hartree-Fock single particle basis. We study Coulomb effects 
as well as the effect of BCS and quasiparticle random phase 
approximation (QRPA) correlations. A measure of isospin mixing in the 
approximate ground state is defined by means of the expectation value 
of the isospin operator squared in $N=Z$ nuclei (which is generalized 
to $N\ne Z$ nuclei). Starting from strict Hartree-Fock approach without 
Coulomb, it is shown that the isospin breaking is negligible, on the 
order of a few per thousand for $(N-Z)=6$, increasing to a few percent 
with Coulomb. Pairing correlations induce rather large isospin mixing 
and Fermi transitions of the forbidden type ($\beta ^-$ for $N\le Z$, 
and $\beta ^+$ for $N\ge Z$). The enhancement produced by BCS 
correlations is compensated to a large extent by QRPA correlations 
induced by isospin conserving residual interactions that tend to 
restore isospin symmetry.

\end{abstract}


\section{Introduction}

It is well known\cite{rs} that the selfconsistent mean field hamiltonian
breaks the symmetries of the exact hamiltonian. The best known example is
the breaking of rotational invariance by the selfconsistent mean field of
deformed nuclei. Rotational invariance breaking often leads to ground states
of Hartree-Fock (HF) or Hartree-Fock-Bogoliubov (HFB) type with large
expectation values of the squared angular momentum operator. For heavy
well deformed even-even nuclei one typically has\cite{moya,j2}:

\begin{equation}
\langle J_{\perp }^2\rangle \gtrsim 100
\end{equation}
where $J_\perp $ is the angular momentum operator component perpendicular
to the symmetry axis and $\langle \; \rangle $ means expectation value
in the HF or HFB ground state. In these instances the HF or HFB wave
function can be interpreted as a wave packet from which the ground state
rotational band can be obtained by angular momentum projection. As a matter
of fact, the angular momentum projection can be carried out through an
expansion in powers of $1/\langle J_\perp ^2\rangle $\cite{villars} which,
to lowest order, provides a factorization approximation formally identical
to that of Bohr and Mottelson \cite{bohrmot}.

Unlike rotational invariance in ordinary space, rotational invariance in
isospin space is not an exact symmetry of the actual total nuclear
hamiltonian. The Coulomb force is non-isoscalar and thus, the actual nuclear
states may have isospin mixing. Isospin mixing in the ground state may allow
for Fermi (F) transitions in $N=Z$ nuclei, and it is a point of present debate 
because of its implications on parity violation experiments on $^{12}$C 
\cite{donnelly}, as well as in the analysis of superallowed Fermi transitions 
as a test of the unitarity of the Cabbibo-Kobayashi-Maskawa matrix \cite{towner}.

On the other hand, even with isoscalar nuclear hamiltonians, the
selfconsistent mean field may break isospin invariance, and particularly so
the selfconsistent quasiparticle mean field (HF+BCS or HFB). Therefore in
studying Fermi transitions and isospin mixing in the nuclear ground state,
it is important to know to what extent the theoretical results respond to
realistic properties of the interactions used in the calculations, or rather
to spurious mean field contributions.

It is important in many respects to know the value of the quantity

\begin{equation}
\langle T_\perp ^2\rangle =\langle T^2\rangle -\left( \frac{N-Z}{2}
\right) ^2 \, .
\end{equation}
If $\langle T_\perp ^2\rangle $ is large when $N=Z$ one may consider, in
analogy to the case $\langle J_\perp ^2\rangle \gg 1$, that the mean
field ground state is a superposition of several $T-$eigenstates. In such a
case, one may generate a corresponding isospin rotational band by isospin
projection \cite{raduta}, in analogy to the above mentioned angular momentum 
projection method, or one may even use an isospin-cranked mean field approximation.
The latter method has been discussed by Wyss et al. \cite{wyss} in a somewhat
different context. On the contrary, if $\langle T_\perp ^{2}\rangle $ is
small ($\langle T_\perp ^{2}\rangle \lesssim 1$ for $N=Z$), it means that 
isospin mixing is small. In this case the mean field ground state is nearly 
an eigenstate of total isospin, no isospin projection may be required, and 
the influence of isospin mixing forces can be reliably studied.

In the shell model context \cite{zuker} large isospin mixing forces (or matrix
elements) have been considered. In this paper we restrict our consideration 
to the Coulomb force as treated in standard mean field 
calculations \cite{vautherin}.

We study Fermi transitions and isospin properties of ground states of
several nuclei, from stable to proton rich isotopes, at various levels
of approximation. We first consider mean field ground states with and
without pairing correlations and with and without isospin breaking
interactions (Coulomb force). Next we take into account isospin dependent
residual interactions and consider QRPA-correlated ground states. The isospin
restoring effect of QRPA-correlations due to isospin conserving residual
interactions is discussed.

Other HF, Tamm-Dancoff approximation (TDA) or RPA studies of isospin impurities
in the ground states of several $N=Z$ nuclei, as well as studies of the effect
of such impurities on superallowed Fermi and Gamow-Teller (GT) $\beta$-decay 
can be found in Refs.\cite{auerbach,hamamoto} and references therein. 
The latter works show that a simple perturbative
treatment of the Coulomb interaction using the same unperturbed wave functions
for neutrons and protons will lead to an overestimate of isospin mixing and
that the use of selfconsistent solutions is essential in calculating the values 
of isospin mixing probabilities. The consistency between the interaction
producing the single particle spectrum and the strength of the residual 
particle-hole interaction is an important ingredient in our calculations.
This consistency has been shown to be essential in many RPA calculations
of giant resonances \cite{auerbach,bertsch,raduta1}, as well as in double
beta-decay \cite{raduta2}.

The paper is organized as follows: In Sec. II we present results on Fermi 
strength distributions. In Sec. III we compare $\langle T_\perp ^{2}\rangle $
values to $\langle J_\perp ^{2}\rangle $ values in the mean field
approach and we discuss $\langle T_\perp ^{2}\rangle $
values in different approaches. In Sec. IV we summarize the main conclusions.

\section{ Fermi Strength Distributions}

In previous publications we studied $\beta $-decay strength functions 
\cite{beta1,beta2} of several isotopic chains in the $A\simeq 60-80$ region, 
within the mean field context and beyond. We performed QRPA calculations on 
top of a quasiparticle basis obtained from a selfconsistent deformed Hartree-Fock
approach with density-dependent Skyrme forces. In this work we use the Skyrme
force SG2 \cite{giai} as a representative of these forces. We use Gamow-Teller
and Fermi residual interactions consistent with the mean field
single particle basis, both being derived from the same two-body Skyrme
interactions. In those works attention was focused on Gamow-Teller strength
distributions which are dominant (see Fig.~1). At variance with standard 
shell-model calculations \cite{shell}, in our selfconsistent mean field 
based calculations \cite{beta1,beta2}, isospin is not an exact quantum number, 
and one may wonder how large can be the effect of isospin breaking. We
investigate this question here focusing on Fermi transitions where isospin
breaking effects can be expected to be of most importance.

The Fermi strength distribution can be written as

\begin{equation}
B^{F^{\pm }}(E)=\sum _f\delta (E-E_f)B_f^{F^{\pm }}\, ,
\label{bf}
\end{equation}
with

\begin{equation}
B_f^{F^{\pm }}=\left| F_f^{\pm }\right| ^2\, ;\qquad
F_f^{\pm }=\left\langle f\mid T_\pm \mid 0\right\rangle \, ,
\label{bfx}
\end{equation}
where $T_\pm $ are the rising and lowering isospin operators, 
$\left| 0\right\rangle $ represents the ground state and 
$\left| f\right\rangle $ are the proton-neutron (pn) excited states.

For the even-even system both initial and final states are represented
by factorized  wave functions of the Bohr-Mottelson type with a common
collective wave function $D_{00}^0$ and intrinsic wave functions
$\left| 0\right\rangle $ and $\left| f\right\rangle $.
In the deformed mean field approximation $\left| f\right\rangle $
represents a one-particle one-hole or a two-quasiparticle excitation, 
connected to the HF or HFB ground state of the parent nucleus 
by the $T_\pm $ operator,

\begin{equation}
F_f^+=2\sum _{ip\,i'n}\left\langle i'\mid i\right\rangle 
u_{i'n}v_{ip}\langle f\mid \alpha _{i'n}^+\,\alpha _{\bar{\imath}p}^+
\mid 0\rangle \, ,
\label{f-}
\end{equation}

\begin{equation}
F_f^-=2\sum _{ip\,i'n}\left\langle i'\mid i\right\rangle ^{\ast }
v_{i'n}u_{ip}\langle f\mid \alpha _{ip}^+\, \alpha _{\bar{\imath}'n}^+
\mid 0\rangle \, .
\label{f+}
\end{equation}

The single particle states $|i>$ are characterized by the eigenvalues
$\Omega_i$ of $J_z$ and by parity $\pi_i$. They are expanded in terms
of the eigenstates of an axially symmetric harmonic oscillator in 
cylindrical coordinates, given in terms of Hermite and Laguerre polynomials 
\cite{vautherin,beta1}:

\begin{equation}
\left| i\right\rangle = \sum_{N}\frac{(-1)^N+\pi_i}{2}\sum_{n_r,n_z,\Lambda
\geq 0,\Sigma} C^i_{Nn_rn_z\Lambda\Sigma}\left|Nn_rn_z\Lambda\Sigma\right\rangle 
\end{equation}
with $\Omega_i=\Lambda+\Sigma \geq 1/2$. For each $N$, the sum over $n_r,n_z,\Lambda$
is extended to the quantum numbers satisfying $2n_r+n_z+\Lambda=N$. The sum over
$N$ goes from $N=0$ to $N=10$. The overlaps $<i'|i>$ are given by

\begin{equation}
\left\langle i'\mid i\right\rangle = \sum_{Nn_z\Lambda\Sigma}
C^{i'}_{Nn_rn_z\Lambda\Sigma} C^i_{Nn_rn_z\Lambda\Sigma}
\end{equation}

In Eqs. (\ref{f-},\ref{f+}) $ip\,i'n$ indicates a proton, neutron state with 
quantum numbers $i,i'$ with $\Omega_i=\Omega_{i'}>0$ and 
$v_{i\tau }(u_{i\tau })$ the probability amplitude that the single
particle state $i\tau $ be occupied (empty) in the ground state. For each 
2-quasiparticle excitation $i'n,\bar{\imath}p$ the value of the energy is 
$E_{i'n}+E_{\bar{\imath}p}=E_f.$  In the limit of no pairing correlations in
the ground state (i.e., in the HF limit) $v_{i\tau }$ goes to 1 and 0 for
$i\tau $ levels below and above the $\lambda _\tau $ Fermi level, respectively. 
In this limit $E_f=\varepsilon _{i'n}-\varepsilon _{ip}$ for 
$\beta ^+$ and $N< Z$; $E_f=\varepsilon _{ip}-\varepsilon _{i'n}$ 
for $\beta ^-,\, N> Z$. We use the convention $\tau =p=-1/2,\quad
\tau =n=1/2\quad (i.e.,\ t_+\left| p\right\rangle =\left|
n\right\rangle ,t_-\left| n\right\rangle =p).$

In the QRPA method $\left| f\right\rangle $ represents a pn-QRPA phonon
state and $\left| 0\right\rangle $ the QRPA correlated ground state. The
expression for the Fermi strength is then given by

\begin{equation}
F_f^+=2\sum_{ip\,i'n}
\left\langle i'\mid i\right\rangle \left[ u_{i'n}v_{ip}
X_{i'n\,\,ip}^f+v_{i'n}u_{ip}Y_{i'n\,ip}^f\right] \, ,
\label{frpa}
\end{equation}
with 
$X_{i'n\,ip}^f$ and $Y_{i'n\,ip}^f$
the forward and backward amplitudes for the two-quasiparticle
component $i'n\, ip$ of the QRPA mode 
$\left| f\right\rangle =\Gamma _f^+\left| 0\right\rangle ,$\ of energy 
$E_f=\omega _f.$

An analogous expression is obtained for $F_f^-$ changing $X-$amplitudes
by $Y-$amplitudes. The calculations of these amplitudes have been done as
described in ref.\cite{beta1}. We use separable residual interactions as 
those derived in ref.\cite{beta1}. We note that the residual Fermi 
interaction $(\vec{\tau}_1\cdot \vec{\tau}_2)$
(as well as the GT $(\vec{\sigma}_1\cdot \vec{\sigma}_2)\, 
(\vec{\tau}_1\cdot \vec{\tau}_2)$) is an isospin restoring force and largely 
eliminates the spurious contributions due to isospin breaking that may be 
present in mean field calculations.

In Fig.~1 we compare Fermi strengths to GT strengths taking as an example 
$^{74}$Kr. The Fermi strength given by Eq. (\ref{bf}), as well as the 
corresponding expression for the Gamow-Teller strength, have been transformed 
into continuous curves by a folding procedure using $\Gamma =1$ MeV width
Gaussians. The strength distributions are plotted versus the excitation
energy of the daughter nucleus. The dominance of Gamow-Teller strengths over 
Fermi strengths is apparent for both $\beta ^+$ and $\beta ^-$.
One clearly sees that the already small mean field Fermi strengths get
further reduced in RPA. The general reduction of the strengths in QRPA can
be traced back to the isospin restoring role of the residual interactions,
being maximal for the $\beta ^+$ Fermi strength which is strongly reduced.
The results shown in this figure are obtained taking into account the 
Coulomb interaction in the mean field.

In the next figures (Fig.~2 to Fig.~5) we show the role of Coulomb interaction, 
as well as that of pairing and RPA correlation effects, on Fermi strengths
calculated for the Kr isotopic chain for $A$=70 to $A$=74.

In Fig.~2 we show the Fermi strengths obtained in the limit of small pairing 
($\Delta =0.1$ MeV) and no Coulomb force. Starting with $^{70}$Kr on the left,
we observe that the $\beta ^+$strength (plotted in the upper panels) rapidly 
decreases with increasing A while the $\beta ^-$ strength (plotted in the 
bottom panels), rapidly increases with $A$. We also observe that the 
reduction of the strength in going from mean field (MF) to QRPA is much 
stronger in the isospin-forbidden cases.

To clarify what we mean by isospin-forbidden cases let us recall that if the 
ground state $\left| 0\right\rangle $ is the exact ground state of the system, 
in the absence of Coulomb force, it is an eigenstate of the isospin operator
with eigenvalues $T_z=(N-Z)/2$ and $T=\left| T_z\right|$.

Hence, it is clear that in this case

\begin{eqnarray}
T_+ \left| 0\right\rangle &=& T_+ \left| T\, T_z \right\rangle
=\sqrt{(Z-N)}\, \left| T\, T_z +1 \right\rangle \text{ for } N<Z \, ,
\label{t+}   \\
&=&0 \quad \text{ for } N\ge Z \, . \nonumber
\end{eqnarray}

\begin{eqnarray}
T_-\left| 0\right\rangle &=&T_-\left| T\, T_z \right\rangle
=\sqrt{(N-Z)}\left| T\, T_z -1 \right\rangle \text{ for } N>Z \, ,
\label{t-}  \\
&=&0 \quad \text{ for }N\le Z  \, .\nonumber
\end{eqnarray}

So that in Eqs.(\ref{bf},\ref{bfx}) there is a single $\left| f\right\rangle $ 
state $\left| f\right\rangle =\left| T\, T_z\pm 1\right\rangle $ (with 
$E_f=E_0$) which is accessible by the $\beta _+$ (if $N<Z$) or the 
$\beta _-$ (if $N>Z$) operators. Accordingly, we call isospin-forbidden 
transitions to the $\beta ^+$ transitions in Kr isotopes with $A\geq 72$ and 
to the $\beta ^-$ transitions with $A\leq 72$. The fact that all 
isospin-forbidden transitions in Fig.~2 are small indicates that
isospin breaking is small in the mean field approximation and gets further
reduced in QRPA.

As we include the Coulomb interaction, $T$ is no longer an exact quantum
number, and isospin breaking (although very small) becomes an actual property 
of the system. In the presence of the Coulomb interaction Eqs.(\ref{t+},\ref{t-}) 
are no longer strictly valid, allowing for $\beta ^{\pm }$ transitions in the
isospin-forbidden cases. When we compare the results in Fig.~3, which we obtained
including the Coulomb force, with the corresponding ones in Fig.~2 we observe
an increase of the $\beta ^+$ $^{72}$Kr, $^{74}$Kr strengths,  as well as of 
the $\beta ^-$ $^{70}$Kr, $^{72}$Kr strengths. This small increase of the 
isospin-forbidden transitions of Fig.~2 is solely due to the effect of the
Coulomb force on Fermi transition and it is a signature of the corresponding
small isospin breaking.

One may also observe that in the absence of Coulomb force (Fig.~2) the 
$\beta ^{+}$ and $\beta ^-$ strength distributions are identical in $^{72}$Kr,
but the two distributions become somewhat different when the Coulomb force
is taken into account (Fig.~3).

Finally in Fig.~4 we show the $\beta ^+$ and $\beta ^-$ strengths for
the same nuclei obtained with realistic pairing gaps ($\Delta \sim 1.5$
MeV, see Table 2). For the allowed transitions ($\beta ^+$ in $^{70}$Kr, 
$\beta ^-$ in $^{74}$Kr) one can see by comparison with Fig.~2 that the 
main effect of pairing is to open up more transition channels and to 
moderately reduce the strengths of the dominant peaks. More dramatic is
the effect of pairing on the isospin-forbidden transitions which grow up by 
orders of magnitude in all cases, even though they remain considerably smaller 
than the allowed ones. This growth of the isospin-forbidden strengths is mainly
connected to the stronger isospin breaking of the quasiparticle mean field
with increasing pairing gaps. It is present irrespective of whether Coulomb
interaction is included or not in the calculation (see also Fig.~5). One also
sees that the QRPA results strongly reduce the strengths of the
isospin-forbidden $\beta ^{\pm }$ transitions as compared to their mean field
values.

\section{Isospin mixing and expectation values of isospin operators}

As mentioned in the Introduction the amount of angular momentum mixing in
the mean field ground state of axially symmetric deformed nuclei is measured
by the expectation value of the squared angular momentum operator
perpendicular to the symmetry axis z.

\begin{equation}
\left\langle J_{\perp }^2\right\rangle \equiv \left\langle
J^2\right\rangle -\left\langle J_z\right\rangle ^2=\frac{1}{2}
\sum_f\left| \left\langle f\right| J_+\left| 0\right\rangle \right|
^2+\left| \left\langle f\right| J_-\left| 0\right\rangle \right| ^2 \, .
\end{equation}
Similarly, taking the z axis in isospin space in the standard way 
($\hat{T}_z=(\hat{N}-\hat{Z})/2$, 
or $T_z=(N-Z)/2$), we can measure the amount of isospin mixing by the
expectation value of  $T_{\perp }^2=T_x^2+T_y^2$, or

\begin{equation}
\left\langle T_{\perp }^2\right\rangle =\frac{1}{2}\sum_f\left(\left|
\left\langle f\left| T_+\right| 0\right\rangle
\right| ^2+\left| \left\langle f\left| T_-\right| 0\right\rangle
\right| ^2\right) \, .
\end{equation}
Using the definition of Fermi transition amplitudes in Eqs.(\ref{f-},\ref{f+}) 
one immediately gets the identity

\begin{equation}
\langle T_{\perp }^2\rangle =\frac{1}{2}\left( S_{F^+}+S_{F^-}\right) \, ,
\end{equation}
with

\begin{equation}
S_{F^{\pm }}=\sum_f\left| F_{f}^{\pm }\right|^2 =\sum_f B_f^{F^{\pm }}\, .
\label{sf+-}
\end{equation}

In Tables 1 and 2 we compare $\left\langle T_{\perp }^2\right\rangle $
values to $\left\langle J_{\perp }^2\right\rangle $ values. The
$\left\langle J_{\perp }^2\right\rangle $ values are calculated \cite{moya}
using the expression

\begin{equation}
\left\langle J_{\perp }^2\right\rangle =\sum_{\tau =p,n}\, \sum_{i\tau
\,k\tau }\left( u_{i\tau }v_{k\tau }-u_{k\tau }v_{i\tau }\right) ^2\left[
\left| \left\langle i_{\tau }\left| j_+\right| k_{\tau }\right\rangle
\right| ^2+\frac{1}{2}\left| \left\langle i_{\tau }\left| j_+\right| 
\bar{k}_{\tau }\right\rangle \right| ^2\right] \, .
\end{equation}

Table 1 corresponds to calculations in the no pairing limit ($\Delta =0)$.
Table 2 contains the results obtained taking into account BCS pairing
correlations with realistic values of the pairing gaps (listed in the
tables).

To see the amount of isospin mixing introduced by the Coulomb force we give
$\left\langle T_{\perp }^2\right\rangle $ values obtained with and without
Coulomb force.

As seen in the tables, the values of $\left\langle J_{\perp}^2\right\rangle $ 
are large for the deformed isotopes. They decrease with decreasing deformation 
and become 0 ($\left\langle J_{\perp}^2\right\rangle =0)$ for spherical 
isotopes ($^{76}$Kr, $^{78}$Kr).

Comparing the $\left\langle J_{\perp }^2\right\rangle $ values in Tables
1 and 2, one can see that $\left\langle J_{\perp }^2\right\rangle $
decreases with pairing. In the no pairing limit for the prolate shape of 
$^{74}$Kr, where the deformation value reaches the value $\beta \simeq 0.4$,
one gets $\left\langle J_{\perp }^2\right\rangle \simeq 90.$ Pairing
correlations with realistic gap values reduce this large $\left\langle
J_{\perp }^2\right\rangle $ value by one third.

This is in contrast to $\left\langle T_{\perp }^2\right\rangle $ values
which are increased when pairing correlations are included. The large 
$\left\langle J_{\perp }^2\right\rangle $ values for the well deformed
shapes are in contrast with the small values of 
$\left\langle T_{\perp}^2\right\rangle _0$, with which they are to be 
compared. The $\left\langle T_{\perp }^2\right\rangle _0$ values, 
given in the tables along with $\left\langle T_{\perp }^2\right\rangle $ 
values, are defined as

\begin{equation}
\left\langle T_{\perp }^2\right\rangle _0=\left\langle T_{\perp}^2
\right\rangle -\left| \frac{N-Z}{2}\right| \, .
\label{t20}
\end{equation}

To clarify the meaning of this magnitude, we recall that in the limit in
which the ground state is an isospin eigenstate, with
$T=\left| T_z\right| =\left| (N-Z)/2\right| $, we have that

\begin{equation}
\left\langle T_{\perp }^2\right\rangle =\left[ T (T+1)
-T_z^2\right] =\left| T_z\right| =\left| \frac{N-Z}{2}\right| \, .
\end{equation}

Therefore $\left\langle T_{\perp }^2\right\rangle _0$ gives a better
measure of the isospin mixing than $\left\langle T_{\perp }^2\right\rangle $ 
in isotopes with $\left| N-Z\right| \neq 0$.

We see in Table 1 that without pairing and Coulomb 
$\left\langle T_{\perp}^2\right\rangle =
\left\langle T_{\perp}^2\right\rangle _0 =0$ in $N=Z$ isotopes, i.e., 
there is no isospin mixing in HF approach when isospin conserving two-body 
force is used. The amount of isospin mixing introduced by the inclusion of 
the Coulomb force in the Hartree-Fock approximation can be seen comparing 
the $\left\langle T_{\perp }^2\right\rangle _0$ values with and without 
Coulomb in Table 1. For the $N\ne Z$ isotopes in the HF approach the amount 
of isospin mixing as measured by $\left\langle T_{\perp }^2\right\rangle _0$ 
is on the order of a few percent with no Coulomb and gets somewhat increased 
when Coulomb force is included. It is interesting to see that the 
$\left\langle T_{\perp }^2\right\rangle _0$ value tends to decrease with 
increasing $\left| T_z\right| =\left| (N-Z)/2 \right| $ value.
This is in agreement with spectroscopic observations and confirms the idea
that the larger is the value of $\left| N-Z\right| $ the better is the
isospin quantum number.

As already mentioned, pairing correlations increase the isospin mixing,
which as seen in Table 2 reaches a maximum in the $N=Z$ $^{72}$Kr isotope
where $\left\langle T_{\perp }^2\right\rangle _0\simeq 2.2$.

Yet if we compare this latter value with the 
$\left\langle J_{\perp}^2\right\rangle $ 
value ( $\left\langle J_{\perp }^2\right\rangle \sim 32)$ for the same nucleus, 
we see that one cannot really talk of an isospin rotational band.
Indeed if we write the HF+BCS state $\phi $\ of $^{72}$Kr as a linear
combination of isospin eigenstates $\phi _T$\

\begin{equation}
\phi =\sum_T\ C_T\ \phi _T\, ,\qquad \sum _T \left| C_T\right| ^2=1 \, ,
\end{equation}
one sees that the mixing of $T\neq 0$ components is quite small.
Indeed, $\left\langle T_{\perp }^2\right\rangle \simeq 2$ means that

\begin{equation}
\left\langle T_{\perp }^2\right\rangle =6\left| C_2\right| ^2+
20\left| C_4\right| ^2+.......\simeq 2 \, ,
\end{equation}
where one sees that $\sum_{T>0} \left| C_T\right| ^2\lesssim 0.3$ and 
$\left| C_{2(n+1)}\right| ^{2}<\left| C_{2n}\right| ^2$.

A corresponding expansion of $\phi $ in orthonormalized angular
momentum eigenfunctions

\begin{equation}
\phi =\sum_J A_J\varphi _J\, ,\qquad  \sum_J \left| A_J\right| ^2=1 \, ,
\end{equation}
gives

\begin{eqnarray}
\left\langle J_{\perp }^2\right\rangle =6\left| A_2\right| ^2 +
20\left| A_4\right| ^2+ ... &\simeq  & 32 \text{ for }\Delta =1.5 \text{ MeV}\, ,  
\nonumber \\
&\simeq & 63 \text{ for }\Delta =0 \, ,
\end{eqnarray}
which provides no constraint on the amount of $J=2,4$ angular momentum
components. This exercise shows that while angular momentum projection
can be used to build up a rotational band, isospin projection can hardly
be used to construct an isospin rotational band out of $\phi$.

Let us now consider what happens when we calculate the expectation values of
$T_{\perp }^2$ in the pn-QRPA correlated ground state. The results
obtained when we take into account QRPA correlations induced by Fermi
residual interactions of the form $\chi _F\ (\vec{\tau}_1\cdot \vec{\tau}_2)$
can be seen in Tables 3 and 4. In these tables we give the values obtained
for the total strengths $S_{F^+}$ and $S_{F^-}$ together with 
$\left\langle T_{\perp }^2\right\rangle _0$ with and without Coulomb
interaction.

In the QRPA approximation the total strengths for $\beta ^+$ and $\beta ^-$ 
Fermi transitions are given by Eq.(\ref{sf+-}), where the sum over $f$ runs 
over all the pn-QRPA solutions and the 
$\beta ^{\pm }$ amplitudes $F_f^{\pm }$ are given by Eq.(\ref{frpa}).

The QRPA solutions \cite{beta1} satisfy the orthonormalization conditions

\begin{eqnarray}
2\sum_f \left( X_{i'n\,ip}^f\; X_{j'n\,jp}^{f\,\ast }-
Y_{i'n\,ip}^f\; Y_{j'n\,jp}^{f\,\ast }\right) &=&
\delta _{ij}\,\delta _{i'j'} \, ,\\
\sum_f\left( X_{i'n\,ip}^f\; Y_{j'n\,jp}^{f\,\ast }-
Y_{i'n\,ip}^f\; X_{j'n\,jp}^{f\,\ast }\right) &=&0 \, .
\end{eqnarray}
Using these orthonormalization conditions one can show that

\begin{eqnarray}
S_{F^+} &=&Z-2\sum_{ii'}\left\langle i'\mid i\right\rangle ^{\ast }
\left\{ \left\langle i'\mid i \right\rangle \, v_{i'n}^2\,
v_{ip}^2-\sum_f\left( Y_{i'n\,ip}^f\right) 
\left[ v_{i'n}u_{ip}F_f^+
+u_{i'n}v_{ip}F_f^-\right] \right\} \, , 
\label{sf+}   \\
S_{F^-} &=&N-2\sum_{ii'}\left\langle i'\mid i\right\rangle ^{\ast }
\left\{ \left\langle i'\mid i\right\rangle \, v_{i'n}^2\,
v_{ip}^2-\sum_f\left( Y_{i'n\,ip}^f\right) 
\left[ v_{i'n}u_{ip}F_f^+ 
+u_{i'n} v_{ip}F_f^-\right] \right\} \, ,
\label{sf-}
\end{eqnarray}
with $F_f^{\pm }$ as given in Eq.(\ref{frpa}). 

These equations explicitly show
that, as pointed out in ref.\cite{split}, the total $\beta ^+ (\beta ^-)$
strength splits into a one-body term that counts the total number of protons
(neutrons), and a two-body term that depends on the two-body correlations
included in the descriptions of the nuclear ground state. The two-body term
is identical in $F^+$ and $F^-$ summed strengths, and as a result one
gets the Ikeda sum rule

\begin{equation}
S_{F^-}-S_{F^+}=N-Z \, ,
\end{equation}
which is always satisfied in our calculations.

The last term in Eqs.(\ref{sf+},\ref{sf-}) is the QRPA correlation term

\begin{equation}
C_{QRPA}=2\sum_f \sum_{ip\,i'n} \left( Y_{ipi'n}^f\right) 
\left\langle i' \mid i\right\rangle ^{\ast }
\left[ v_{i'n}u_{ip} F_f^+
+u_{i'n} v_{ip}F_f^-\right] \, .
\label{crpa}
\end{equation}
This term is zero when there are no QRPA correlations. If the strength of
the residual interaction is zero ($\chi _F=0)$\ the $Y-$amplitudes are zero.

The second term in Eqs.(\ref{sf+},\ref{sf-}) contains the pure pairing 
correlation and it reaches its minimum value in the limit of no pairing and 
no Coulomb force. Indeed in the HF limit one has that

\begin{equation}
\left( -2\sum_{i\,i'}\left| \left\langle i'\mid i\right\rangle \right| ^2\,
v_{i'n}^2 v_{ip}^2\right) _{HF} =
-2\sum_{i<\lambda_p,\, i'<\lambda_n} \left| \left\langle 
i'\mid i\right\rangle \right| ^2 \, .
\end{equation}
If, in addition, there is no Coulomb force

\begin{equation}
\left\langle i'\mid i\right\rangle =\delta _{i'i}\quad {\rm for }\; N=Z\, ,
\end{equation}
and we have that

\begin{equation}
\left( -2\sum_{i\,i'}\left| \left\langle i'\mid 
i\right\rangle \right| ^2\,v_{i'n}^2 v_{ip}^2 \right) _{HF}=
-2\sum_{i<\lambda_p,\, i'<\lambda_n}\delta _{i'i}
=-Z \quad {\rm for }\; N=Z\, .
\end{equation}
We therefore find that in the limit of no pairing and no Coulomb force:

\begin{equation}
\left\langle T_{\perp }^2\right\rangle =\left\langle T_{\perp}^2
\right\rangle _{HF}=0 \quad {\rm for }\; N=Z\, ,
\label{t21}
\end{equation}
as we found numerically in Table 1. In the general case we can write

\begin{equation}
\left\langle T_{\perp }^2\right\rangle =\frac{1}{2}\left(
S_{F^+}+S_{F^-}\right) =  \left| \frac{N-Z}{2}\right|
+C_{BCS}+C_{QRPA} \, ,
\label{t22}
\end{equation}
with $C_{QRPA}$ the QRPA correlation term defined in Eq.(\ref{crpa}) and 
$C_{BCS} $ the BCS correlation term defined as

\begin{equation}
C_{BCS}=-2\sum_{i'n\,ip}\left| \left\langle i'
\mid i\right\rangle \right| ^2\,v_{i'n}^2 v_{ip}^2+\min (Z,N) \, .
\label{c_bcs}
\end{equation}
Owing to Eq.(\ref{t22}) we see that the value of 
$\left\langle T_{\perp }^2\right\rangle _0$, as defined in Eq.(\ref{t20}), 
is purely due to the correlation terms in Eqs.(\ref{crpa}-\ref{c_bcs}),

\begin{equation}
\left\langle T_{\perp }^2\right\rangle _0=\left\langle T_{\perp}^2
\right\rangle -\left| \frac{N-Z}{2}\right| =C_{BCS}+C_{QRPA}\, ,
\label{tfin}
\end{equation}
and that for $N=Z$ and no Coulomb force 

\begin{equation}
\left\langle T_{\perp }^2\right\rangle =\left\langle T_{\perp}^2
\right\rangle _0 =\left\langle T_{\perp }^2\right\rangle _{HF}\, .
\end{equation}

The results in Tables 1 and 2 show that the BCS correlations increase always
the $\left\langle T_{\perp }^2\right\rangle $ value from its limiting
Hartree-Fock value. On the other hand if we compare the results of Tables 1
and 2 to the results of Tables 3 and 4, which include both correlation terms
(see Eq.(\ref{tfin})), we see that the increase of 
$\left\langle T_{\perp}^2\right\rangle _0$ due to the BCS correlations is 
largely reduced by
the QRPA correlations, which tend to restore isospin invariance.

\section{Summary and concluding remarks}

We have studied isospin mixing properties in several Kr isotopes 
around $N=Z$ and  analyzed their Fermi transitions at various levels 
of approximation.

We have considered first, selfconsistent deformed Skyrme HF mean 
fields with and without Coulomb and pairing interactions. Then, we 
take into account isospin dependent residual interactions and consider 
QRPA correlated ground states. 
We have studied the effect on the Fermi strength distributions of 
isospin breaking interactions (Coulomb force and pairing), as well as 
the effect of QRPA correlations including Fermi type residual interactions,
whose particle-hole strengths are consistently fixed with the Skyrme force.

Taking as a reference the case of a selfconsistent mean field calculation, 
we have seen that in the absence of Coulomb interactions (and in the limit 
of small pairing correlations) the isospin-forbidden transitions 
($\beta ^-$ in $N\le Z$ and $\beta ^+$ in $N\ge Z$) are 
negligible. When the isospin breaking Coulomb interaction is switched on, 
there is an increase of isospin-forbidden Fermi transitions. Although this 
increase is small, it is a signature of isospin breaking. Pairing 
correlations increase by orders of magnitude the isospin-forbidden Fermi 
transitions, a fact that is related to the isospin breaking nature of 
the quasiparticle mean field, which increases with increasing pairing 
gaps. On the other hand, the isospin breaking effects and forbidden Fermi
transitions are reduced when RPA correlations are taken into account.

In analogy with the quantity $\langle J_{\perp }^2\rangle$, 
which measures the amount of angular 
momentum mixing, we have introduced the quantity 
$\langle T_{\perp }^2\rangle _0$  as a measure of the
isospin mixing in the ground state.

In the extreme case of mean field approach without Coulomb interaction
and without pairing correlations, we have 
$\langle T_{\perp }^2\rangle _0=0$ for $N=Z$. There is no isospin 
mixing in HF approach when isospin conserving two-body forces are used.
The amount of isospin mixing introduced by the Coulomb force is small
and the maximum mixing occurs when $N=Z$. In contrast to 
$\langle J_{\perp }^2\rangle$, which 
decreases with increasing pairing, the amount of isospin mixing 
$\langle T_{\perp }^2\rangle _0$ increases with pairing correlations 
and it is also maximum for $N=Z$. The lowest isospin mixing, as measured by 
$\langle T_{\perp }^2\rangle _0$, occurs in HF approximation. Pairing 
correlations increase the mixing and QRPA correlations reduce it.

One may wonder whether, in the deformed nuclei, the calculated Fermi strengths
may contain spurious contributions from higher angular momentum components
in the initial and final wave functions. As mentioned before, the Fermi
strengths are calculated in the laboratory frame in the factorization
approximation of Bohr and Mottelson \cite{bohrmot}.
Using angular momentum projection techniques \cite{moya}, we find that an upper 
bound to such contributions is proportional to $1/<J_{\perp}^2>^2$, where the
values of $<J_{\perp}^2>$ can be found in Tables 1 and 2. Thus, exact angular 
momentum projection in the deformed cases would lead in all cases to less 
than $1\%$ effect in the Fermi strengths.

It is also shown that the total $\beta ^{\pm}$ Fermi strengths can be 
separated into a one-body term 
counting basically the number of nucleons of a given type, and a two-body 
term that depends on the two-body correlations (BCS and RPA). The two-body 
term is identical in both $\beta^+$ and $\beta^-$ summed strengths and the 
Ikeda sum rule is fulfilled as a result of their cancellation. The isospin 
mixing $\langle T_{\perp }^2\rangle _0$ is purely due to the net effect of 
the correlation terms.

\begin{center}
{\Large \bf Acknowledgments} 
\end{center}
This work was supported by Ministerio de Educaci\'on y Ciencia (Spain) under 
contract number BFM2002-03562. One of us (R.A.-R.) thanks Ministerio de 
Educaci\'on y Ciencia (Spain) for financial support. 

%\newpage

\newpage

\begin{center}

\begin{table}[t]
\caption{Comparison of angular momentum and isospin mixing in the mean field
approximation without pairing correlations (HF). The HF value of the 
$\beta-$ deformation parameter is given in the second column.}

\bigskip

\begin{tabular}{ccccccccc}
& $\beta $ & $\left\langle J_{\perp }^2\right\rangle $ & 
\multicolumn{2}{c}{$\left\langle T_{\perp }^2\right\rangle _0$} 
&& \multicolumn{2}{c}{$\left\langle T_{\perp}^2\right\rangle $} 
& IKEDA \\ \cline{4-5} \cline{6-8}
&  &  & Coulomb & No Coulomb && Coulomb & No Coulomb &  \\ \hline
$^{70}$Kr & -0.27 & 50.2 & 0.04 & 0.00 && 1.04 & 1.00 & -2.00 \\ 
$^{72}$Kr & -0.29 & 63.1 & 0.05 & 0.00 && 0.05 & 0.00 &  0.00 \\ 
$^{74}$Kr & -0.15 & 21.1 & 0.05 & 0.01 && 1.05 & 1.01 &  2.00 \\ 
          &  0.39 & 89.9 & 0.06 & 0.02 && 1.06 & 1.02 &  2.00 \\ 
$^{76}$Kr &  0.00 &  0.0 & 0.03 & 0.01 && 2.03 & 2.01 &  4.00 \\
$^{78}$Kr &  0.00 &  0.0 & 0.03 & 0.01 && 3.03 & 3.01 &  6.00 \\
\end{tabular}
\end{table}

\begin{table}[t]
\caption{Same as Table 1 taking into account pairing correlations (HF+BCS)
with fixed pairing gaps $\Delta _p=\Delta_n =\Delta $
given in the second column.}

\bigskip

\begin{tabular}{cccccccccc}
& $\Delta $ (MeV) & $\beta $ & $\left\langle J_{\perp}^2 \right\rangle $ 
& \multicolumn{2}{c}{$\left\langle T_{\perp }^2\right\rangle _0$} && 
\multicolumn{2}{c}{$\left\langle T_{\perp }^2\right\rangle $} & IKEDA \\ 
\cline{5-6} \cline{8-9}
& &  &  & Coulomb & No Coulomb && Coulomb  & No Coulomb & \\ \hline
$^{70}$Kr & 1.5 & -0.23 & 28.5 & 1.2 & 1.1 && 2.2 & 2.1 & -2.0 \\
$^{72}$Kr & 1.5 & -0.24 & 31.8 & 2.2 & 2.2 && 2.2 & 2.2 &  0.0 \\ 
$^{74}$Kr & 1.5 & -0.15 & 13.0 & 1.5 & 1.4 && 2.5 & 2.4 &  2.0 \\ 
          &     &  0.39 & 59.9 & 1.3 & 1.3 && 2.3 & 2.3 &  2.0 \\ 
$^{76}$Kr & 1.65 & 0.0  & 0.0  & 1.3 & 1.2 && 3.3 & 3.2 &  4.0 \\
$^{78}$Kr & 1.75 & 0.0  & 0.0  & 1.0 & 0.9 && 4.0 & 3.9 &  6.0 \\
\end{tabular}
\end{table}

\begin{table}[t]
\caption{Results of QRPA calculations of total $\beta ^+$ and $\beta ^-$
Fermi strengths of Kr isotopes, with and without Coulomb, in the small
pairing limit ($\Delta =0.1$ MeV is used for all isotopes). Also listed
are the values of $\left\langle T_{\perp }^2\right\rangle _0.$ The values
of the deformation parameter $\beta $\ are as in Tables 1 and 2. The
strengths of the particle-hole and particle-particle residual interactions
are $\chi _{ph}^F=0.5$ MeV and $\kappa _{pp}^F=.03$ MeV, respectively 
(see ref.\protect\cite{beta1}).}

\bigskip

\begin{tabular}{ccccccccc}
Isotope & \multicolumn{2}{c}{$S_{F^-}$} && 
\multicolumn{2}{c}{$S_{F^+}$} && \multicolumn{2}{c}{$\left\langle
T_{\perp }^2\right\rangle _0$} \\ 
\cline{2-3} \cline{5-6} \cline{8-9}
& Coulomb & No Coulomb && Coulomb & No Coulomb && Coulomb & No Coulomb \\ \hline
$^{70}$Kr & 0.05 & 0.00 && 2.05 & 2.00 && 0.05 & 0.00 \\ 
$^{72}$Kr & 0.07 & 0.02 && 0.07 & 0.02 && 0.07 & 0.02 \\ 
$^{74}$Kr & 2.03 & 2.00 && 0.03 & 0.00 && 0.03 & 0.00 \\ 
          & 2.05 & 2.00 && 0.05 & 0.00 && 0.05 & 0.00 \\ 
$^{76}$Kr & 4.02 & 4.00 && 0.02 & 0.00 && 0.02 & 0.00 \\ 
$^{78}$Kr & 6.01 & 6.00 && 0.01 & 0.00 && 0.01 & 0.00 \\ 
\end{tabular}
\end{table}

\begin{table}[t]
\caption{Same as Table 3 but for standard values of the pairing gap 
($\Delta =\Delta _n=\Delta _p\sim 1.5$ MeV, as listed in Table 2).}

\bigskip

\begin{tabular}{ccccccccc}
Isotope & \multicolumn{2}{c}{$S_{F^-}$} && \multicolumn{2}{c}{$S_{F^+}$} && 
\multicolumn{2}{c}{$\left\langle T_{\perp }^2\right\rangle _0$} \\ 
\cline{2-3} \cline{5-6} \cline{8-9}
& Coulomb & No Coulomb && Coulomb & No Coulomb && Coulomb & No Coulomb \\ \hline
$^{70}$Kr & 0.43 & 0.41 && 2.43 & 2.41 && 0.43 & 0.41 \\
$^{72}$Kr & 1.15 & 1.11 && 1.15 & 1.11 && 1.15 & 1.11 \\
$^{74}$Kr & 2.50 & 2.44 && 0.50 & 0.44 && 0.50 & 0.44 \\
          & 2.46 & 2.48 && 0.46 & 0.48 && 0.46 & 0.48 \\
$^{76}$Kr & 4.30 & 4.23 && 0.30 & 0.23 && 0.30 & 0.23 \\
$^{78}$Kr & 6.17 & 6.12 && 0.17 & 0.12 && 0.17 & 0.12 \\
\end{tabular}
\end{table}
\end{center}

\newpage

\begin{center}

\begin{figure}[t]
\psfig{file=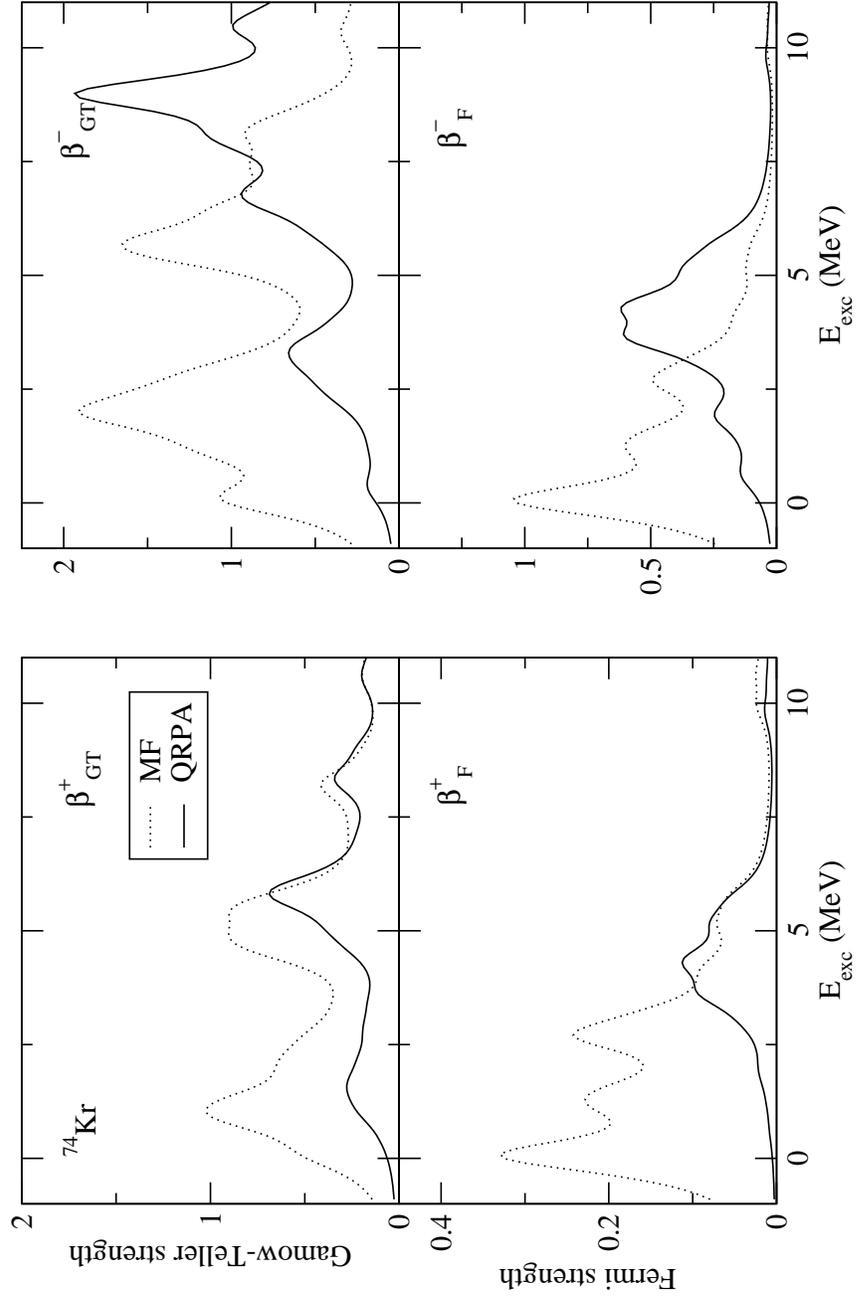,width=0.8\textwidth}
%\vskip 1cm
\caption{Comparison of $\beta ^\pm$ Gamow-Teller and Fermi strength distributions
in both mean field (MF) and QRPA approximations in the case of $^{74}$Kr.}
\end{figure}

\newpage

\begin{figure}[t]
\psfig{file=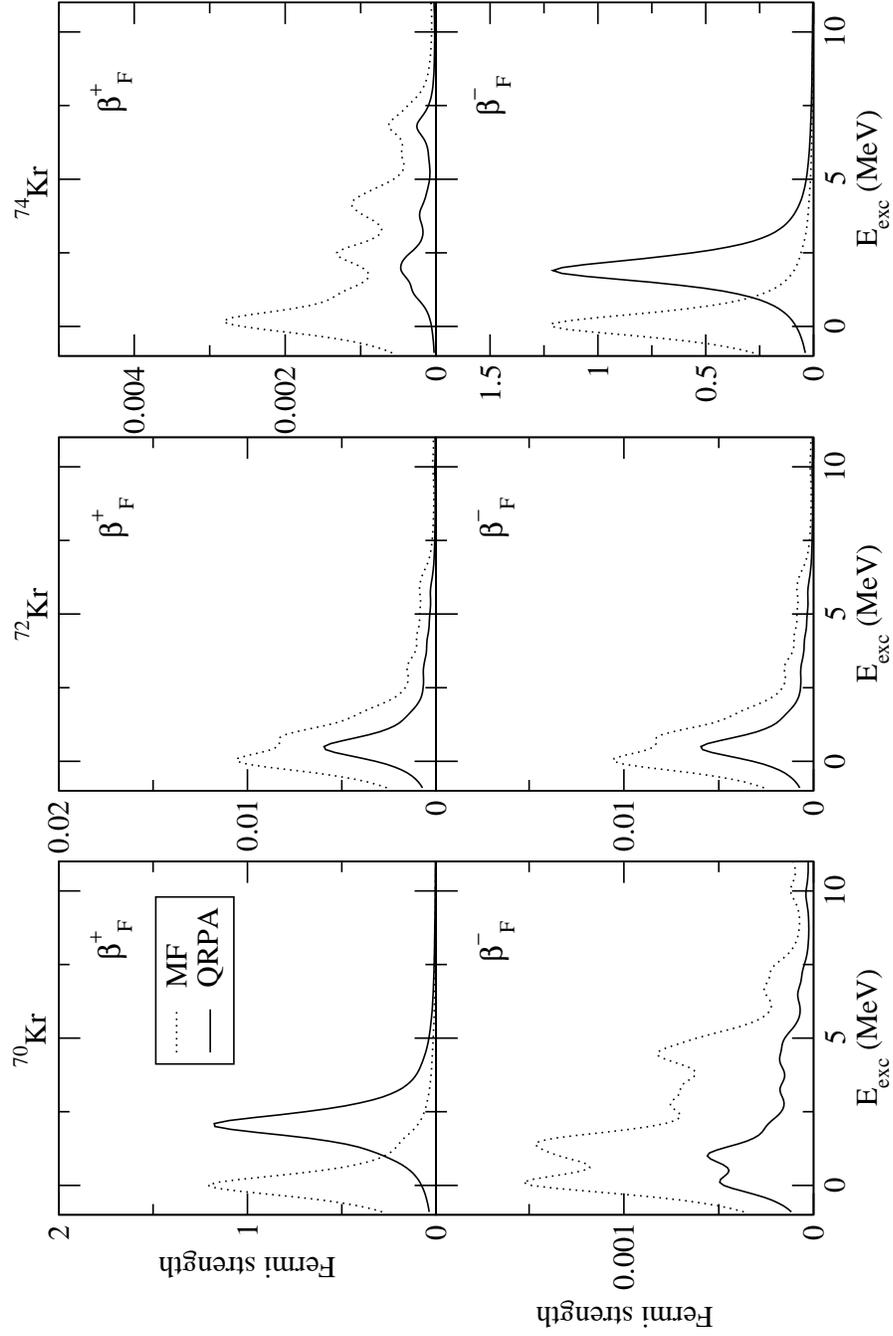,width=0.8\textwidth}
%\vskip 1cm
\caption{Fermi strength distributions $\beta ^\pm$ in $^{70,72,74}$Kr plotted 
as a function of the excitation energy of the daughter nucleus. We compare
results obtained from Skyrme Hartree-Fock calculations (dotted lines) and 
QRPA (solid lines). The results correspond to the case of no Coulomb interaction
and pairing gaps approaching zero ($\Delta=0.1$ MeV).}
\end{figure}

\newpage

\begin{figure}[t]
\psfig{file=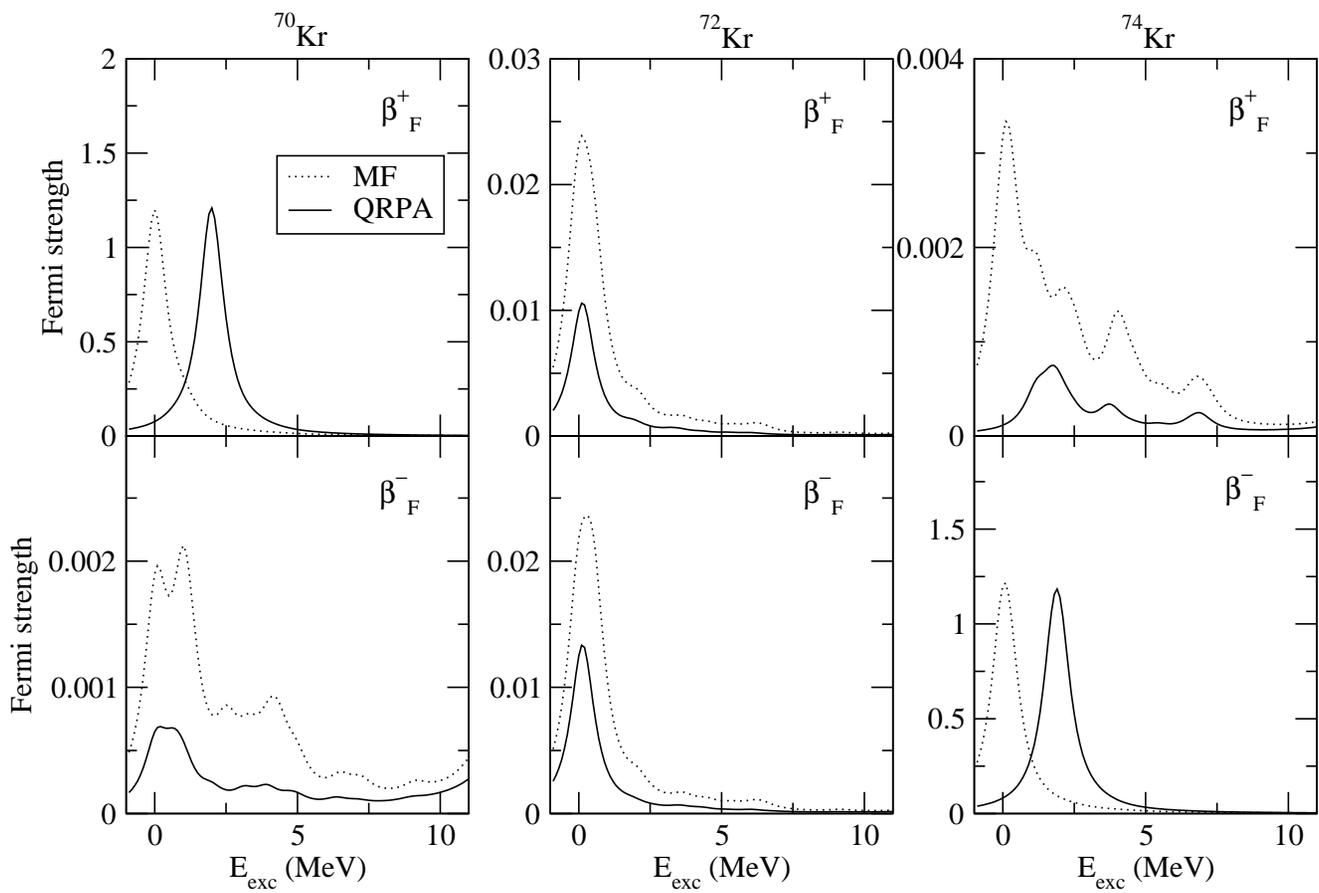,width=0.8\textwidth}
%\vskip 1cm
\caption{Same as in Fig.~2 but with Coulomb interaction.}
\end{figure}

\newpage

\begin{figure}[t]
\psfig{file=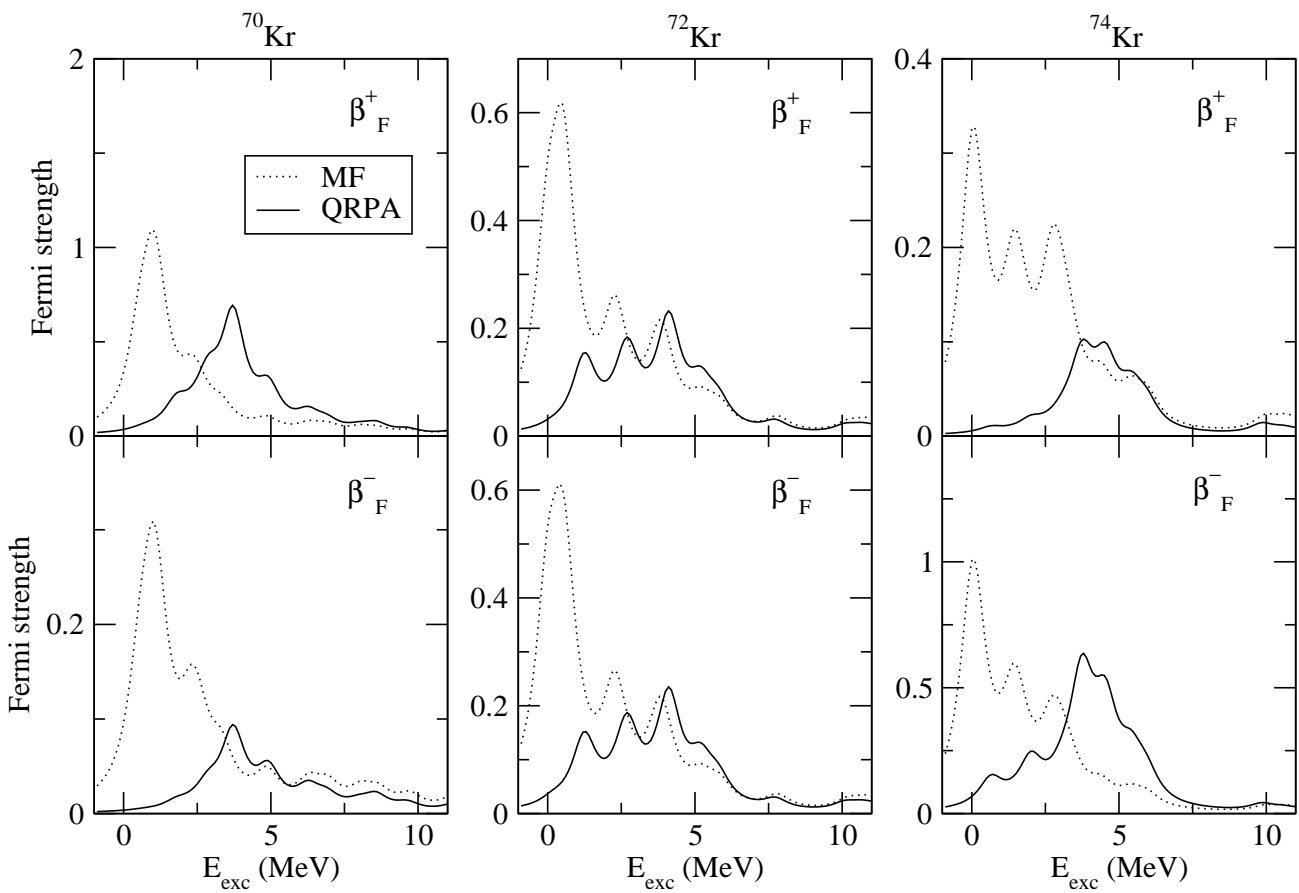,width=0.8\textwidth}
%\vskip 1cm
\caption{Same as in Fig.~2 without Coulomb interaction but including pairing
correlations with realistic gaps (see Table 2).}
\end{figure}

\newpage

\begin{figure}[t]
\psfig{file=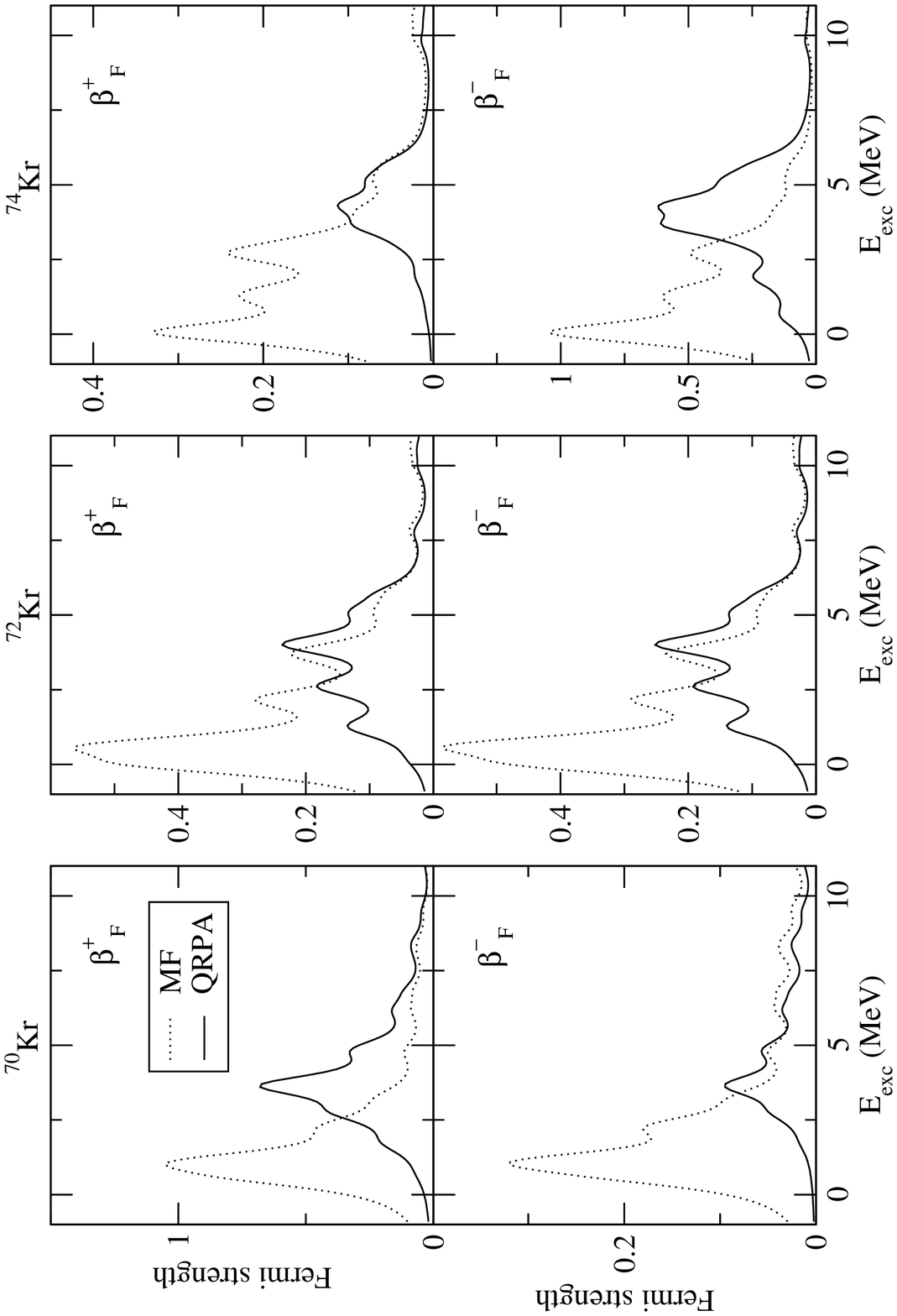,width=0.8\textwidth}
%\vskip 1cm
\caption{Same as in Fig.~4 but including Coulomb interaction.}
\end{figure}

\end{center}


\begin{thebibliography}{00}

\bibitem{rs} A. de Shalit and H. Feshbach, {\em Theoretical Nuclear Physics},
(Wiley, New York, 1974); P. Ring and P. Schuck, {\em The Nuclear Many-body 
Problem}, (Springer-Verlag, New York 1980); I. Talmi,  {\em Simple Models 
of Complex Nuclei}, (Harwood Academic, Switzerland, 1993).

\bibitem{moya} E. Moya de Guerra, Phys. Rep. {\bf 138}, 293 (1986).

\bibitem{j2} H. Flocard, P. Quentin and D. Vautherin, Phys. Lett. B {\bf 46}, 
304 (1973); A. Zarringhalam and J.W. Negele, Nucl. Phys. {\bf A288}, 417 
(1977); E. Moya de Guerra and S. Kowalski, Phys. Rev. C {\bf 20}, 357 (1979).

\bibitem{villars} F. Villars, {\em Many-body Description of Nuclear Structure
and Reactions}, (ed. C. Bloch, Academic Press, New York, 1966).

\bibitem{bohrmot}  A. Bohr and B. Mottelson, {\em Nuclear Structure}, (Benjamin,
New York 1975).

\bibitem{donnelly} M.J. Musolf and T.W. Donnelly, Nucl. Phys. {\bf A546}, 509 
(1992); and T.W. Donnelly, private communication.

\bibitem{towner} I.S. Towner and J.C. Hardy, Phys. Rev. C {\bf 66}, 035501 (2002).

\bibitem{raduta} A.A. Raduta and E. Moya de Guerra, Ann. Phys. (N.Y.) {\bf 284},
134 (2000).

\bibitem{wyss} W. Satula and R.A. Wyss, Acta Phys. Polon. B {\bf 32}, 2441 (2001); 
S. Glowacz, W. Satula and R.A. Wyss, Eur. Phys. J. A {\bf 19}, 33 (2004).

\bibitem{zuker} A.P. Zuker, S.M. Lenzi, G. Mart\'{\i}nez-Pinedo and A. Poves,
Phys. Rev. Lett. {\bf 89}, 142502 (2002).

\bibitem{vautherin}  D. Vautherin and D. M. Brink, Phys. Rev. C {\bf 5} (1972)
626; D. Vautherin, Phys. Rev. C {\bf 7}, 296 (1973); M. Vallieres and D.W.L.
Sprung, Can. J. Phys. {\bf 56}, 1190 (1978).

\bibitem{auerbach} N. Auerbach, Phys. Rep. {\bf 98}, 273 (1983); N. Auerbach
and O.K. Vorov, Phys. Lett. B {\bf 414}, 1 (1997).

\bibitem{hamamoto} I. Hamamoto and H. Sagawa, Phys. Rev. C {\bf 48}, R960 (1993);
J. Dabaczewski and I. Hamamoto, Phys. Lett. B {\bf 345}, 181 (1995); H. Sagawa,
N. Van Giai and T. Suzuki, Phys. Rev. C {\bf 53}, 2163 (1996); G. Colo,
M.A. Nagarajan, P. Van Isacker and A. Vitturi, Phys. Rev. C {\bf 52}, R1175 (1995).

\bibitem{bertsch} G.F. Bertsch and S.F. Tsai, Phys. Rep. {\bf 18}, 126 (1975).

\bibitem{raduta1} A.A. Raduta, L. Pacearescu, V. Baran, P. Sarriguren and E. Moya
de Guerra, Nucl. Phys. {\bf A675}, 503 (2000).

\bibitem{raduta2} A.A. Raduta, P. Sarriguren, A. Faessler and E. Moya de Guerra,
Ann. Phys. (N.Y.) {\bf 294}, 182 (2001).

\bibitem{beta1} P. Sarriguren, E. Moya de Guerra, A. Escuderos, and A.C.
Carrizo, Nucl. Phys. {\bf A635}, 55 (1998).

\bibitem{beta2} P. Sarriguren, E. Moya de Guerra, and A. Escuderos, Nucl. Phys. 
{\bf A658}, 13 (1999); Nucl. Phys. {\bf A691}, 631 (2001); 
Phys. Rev. C {\bf 64}, 064306 (2001).

\bibitem{giai} N. Van Giai and H. Sagawa, Phys. Lett. B {\bf 106}, 379 (1981).

\bibitem{shell} E. Caurier, A. Poves and A.P. Zuker, Phys. Rev. Lett. {\bf 74},
1517 (1995); E. Caurier, G. Martinez-Pinedo, A. Poves and A.P. Zuker, Phys.
Rev. C {\bf 52}, R1736  (1995); P.B. Radha, D.J. Dean, S.E. Koonin, K. Langanke
and P. Vogel, Phys. Rev. C {\bf 56}, 3079 (1997); 
and A. Poves, private communication.

\bibitem{split} E. Moya de Guerra, A.A. Raduta, L. Zamick and P. Sarriguren,
Nucl. Phys. {\bf A727}, 3 (2003).

\end{thebibliography}
\end{document}